\documentclass[english]{article}

\usepackage{spconf}
\usepackage[style=ieee, doi=false, isbn=false, url=false]{biblatex}
\usepackage{hyperref}
\usepackage{bm,bbm}
\usepackage{amsthm,amssymb,amsmath,mathtools}
\usepackage{stmaryrd}
\usepackage{microtype}
\usepackage{algpseudocode,algorithm,algorithmicx}
\usepackage{xcolor,graphicx,tabularx,float}
\usepackage{babel,csquotes}
\usepackage{comment}

\makeatletter

\theoremstyle{plain}
\newtheorem{theroem}{Theorem}

\newtheorem{lemma}[theroem]{Lemma}
\newtheorem{proposition}[theroem]{Proposition}
\newtheorem{corollary}[theroem]{Corollary}

\bmdefine{\bA}{A}
\bmdefine{\ba}{a}
\bmdefine{\bB}{B}
\bmdefine{\bb}{b}
\bmdefine{\bC}{C}
\bmdefine{\bc}{c}
\bmdefine{\bD}{D}
\bmdefine{\bd}{d}
\bmdefine{\bE}{E}
\bmdefine{\be}{e}
\bmdefine{\bF}{F}
\bmdefine{\bG}{G}
\bmdefine{\bg}{g}
\bmdefine{\bH}{H}
\bmdefine{\bh}{h}
\bmdefine{\bI}{I}
\bmdefine{\bi}{i}
\bmdefine{\bJ}{J}
\bmdefine{\bj}{j}
\bmdefine{\bK}{K}
\bmdefine{\bk}{k}
\bmdefine{\bL}{L}
\bmdefine{\bl}{l}
\bmdefine{\bM}{M}
\bmdefine{\bN}{N}
\bmdefine{\bn}{n}
\bmdefine{\bO}{O}
\bmdefine{\bo}{o}
\bmdefine{\bP}{P}
\bmdefine{\bp}{p}
\bmdefine{\bQ}{Q}
\bmdefine{\bq}{q}
\bmdefine{\bR}{R}
\bmdefine{\br}{r}
\bmdefine{\bS}{S}
\bmdefine{\bs}{s}
\bmdefine{\bT}{T}
\bmdefine{\bt}{t}
\bmdefine{\bU}{U}
\bmdefine{\bu}{u}
\bmdefine{\bV}{V}
\bmdefine{\bv}{v}
\bmdefine{\bW}{W}
\bmdefine{\bw}{w}
\bmdefine{\bX}{X}
\bmdefine{\bx}{x}
\bmdefine{\bY}{Y}
\bmdefine{\by}{y}
\bmdefine{\bZ}{Z}
\bmdefine{\bz}{z}

\bmdefine{\balpha}{\alpha}
\bmdefine{\bbeta}{\beta}
\bmdefine{\bgamma}{\gamma}
\bmdefine{\bGamma}{\Gamma}
\bmdefine{\bdelta}{\delta}
\bmdefine{\bDelta}{\Delta}
\bmdefine{\bepsilon}{\epsilon}
\bmdefine{\bvarepsilon}{\varepsilon}
\bmdefine{\bzeta}{\zeta}
\bmdefine{\bmeta}{\eta}
\bmdefine{\btheta}{\theta}
\bmdefine{\bTheta}{\Theta}
\bmdefine{\biota}{\iota}
\bmdefine{\bkappa}{\kappa}
\bmdefine{\blambda}{\lambda}
\bmdefine{\bLambda}{\Lambda}
\bmdefine{\bmu}{\mu}
\bmdefine{\bnu}{\nu}
\bmdefine{\bpi}{\pi}
\bmdefine{\bPi}{\Pi}
\bmdefine{\brho}{\rho}
\bmdefine{\bsigma}{\sigma}
\bmdefine{\bSigma}{\Sigma}
\bmdefine{\btau}{\tau}
\bmdefine{\bupsilon}{\upsilon}
\bmdefine{\bUpsilon}{\Upsilon}
\bmdefine{\bphi}{\phi}
\bmdefine{\bPhi}{\Phi}
\bmdefine{\bchi}{\chi}
\bmdefine{\bpsi}{\psi}
\bmdefine{\bPsi}{\Psi}
\bmdefine{\bomega}{\omega}
\bmdefine{\bOmage}{\Omega}

\newcommand{\cB}{{\mathcal{B}}}

\newcommand{\cN}{\mathcal{N}}

\newcommand{\bbR}{\mathbb{R}}

\newcommand{\sT}{{\mathsf{T}}}

\newcommand{\rE}{\mathrm{E}}

\DeclareMathOperator*{\argmin}{arg\,min}

\DeclareMathOperator{\diag}{diag}

\let\hat\widehat
\let\tilde\widetilde

\newcommand\inp[2]{\left\langle #1,#2 \right\rangle}
\newcommand\norm[1]{\left\lVert #1 \right\rVert}

\newcommand{\ellp}{{\ell^\prime}}

\makeatother

\hypersetup{
    pdftitle={A Framework for Private Communication with Secret Block Structure},
    pdfauthor={Maxime Ferreira da Costa and Urbashi Mitra}
    }

\newcommand*\Let[2]{\State #1 $\gets$ #2}

\title{A Framework for Private Communication with Secret Block Structure}

\name{Maxime Ferreira Da Costa and Urbashi Mitra
	\thanks{This work has been funded in part by one or more of the following: Cisco Foundation 1980393, ONR N00014--15--1--2550, ONR 503400--78050, NSF CCF--1410009, NSF CCF--1817200, NSF CCF--2008927, Swedish Research Council 2018--04359, ARO W911NF1910269, and DOE DE--SC0021417.}
	\thanks{Authors' emails: \texttt{\{{mferreira, ubli}\}@usc.edu}}
	}
	\address{Ming Hsieh Department of Electrical Engineering, Viterbi School of Engineering\\
	University of Southern California, Los Angeles, CA, USA}

\begin{document}

\maketitle

\begin{abstract}
Harnessing a block-sparse prior to recover signals through underdetermined linear measurements has been extensively shown to allow exact recovery in conditions where classical compressed sensing would provably fail. We exploit this result to propose a novel private communication framework where the secrecy is achieved by transmitting instances of an unidentifiable compressed sensing problem over a public channel. The legitimate receiver can attempt to overcome this ill-posedness by leveraging secret knowledge of a  block structure that was used to encode the transmitter's message. We study the privacy guarantees of this communication protocol to a single transmission, and to multiple transmissions without refreshing the shared secret. Additionally, we propose an algorithm for an eavesdropper to learn the block structure via the method of moments and highlight the privacy benefits of this framework through numerical experiments.
\end{abstract}


\section{Introduction}

While security is commonly considered over the \emph{transport layer} and achieved by the mean of cryptographic algorithms, recent advances in \emph{physical layer} security \cite{bloch2011physical}, which aims at exploiting the physical properties of a communication channel to discriminate in favor of legitimate parties, have allowed a leap forward for private communication with modern applications to next-generation wireless systems \cite{poor2017wireless}.
To that end, the compressed sensing framework \cite{donoho2006compressed} has been extensively considers as a mean to ensure privacy~\cite{zhang2016review}. If the sensing matrix is kept secret to an eavesdropper, perfect secrecy can be guaranteed in the information theoretic sense \cite{liang2009information} under restrictive conditions \cite{bianchi2015analysis}. The computational secrecy of this approach have also been discussed~\cite{orsdemir2008security,rachlin2008secrecy}, restricting Eve's ability to recover the encoded message via a polynomial time algorithm.

Motivated by applications to MIMO systems, we focus here instead on a novel model where the sensing matrix (\emph{e.g.} the channel matrix) is imposed by the environment and not left to design by the transmitter. Privacy is rather achieved by sharing  an additional structure on the message with the legitimate receiver, easing the decoding of the message \cite{baraniuk2010model}. On the eavesdropper perspective, the decoding amounts to solving a bilinear inverse problem, which are known to demand much stringent assumptions to be identifiable \cite{choudhary2013identifiability,choudhary2013identifiability2,da2019self,li2015unified}.

\begin{figure}[t]
	\centering
	\includegraphics[width=0.95\columnwidth]{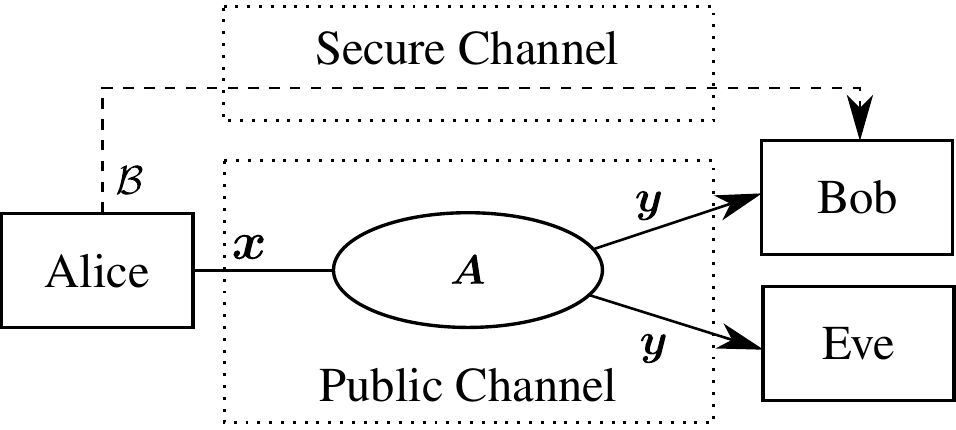}
	\caption{Communication Model with Secure Channel}
	\label{fig:model}
\end{figure}

\subsection{Linear Inverse Problem Based Privacy }
We consider the classical secret communication problem with side information: A transmitter (Alice) wishes to privately transmit a vector $\bx \in \bbR^N$ to a legitimate receiver (Bob) over a public channel. The channel output $\by = f(\bx)$ can be overheard by both Bob and an eavesdropper (Eve). To achieve privacy and prevent Eve from recovering the message $\bx$, Alice and Bob may communicate a low information rate signal over a secure channel that is inaccessible by Eve.

In the proposed setting, the effect of the channel is assumed to be linear and noiseless and modelled by a fat matrix $\bA \in \bbR^{M\times N}$, with $M < N$ so that $\by = \bA \bx$. The matrix $\bA$ is imposed by the environment and is assumed to be known by Bob and Eve. For the purposes of the analysis, we suppose $\bA$ be drawn at random with i.i.d. Gaussian entries $a_{i,j} \sim \cN \left(0, \frac{1}{M} \right)$. Finally, we assume that Eve is aware of the communication protocol established by Alice. The overall communication model is depicted in Fig.~\ref{fig:model}.

In order to ensuring privacy, Alice, who designs the message $\bx$ and the side information, needs to certify two things. First, Bob must be able to provably recover $\bx$ from the observation $\by$ via the side information from the secure channel. Second, Eve cannot provably recover $\bx$ without knowing the side information. Thus, Alice is left to design an inverse problem that is \emph{identifiable} to Bob, but \emph{unidentifiable} to Eve. These goals can typically be achieved by imposing an additional structure on $\bx$, and by privately sharing this structure over the secure channel. For practicality, this structure must be comprised of a small number of bits, and reusable.

\subsection{Contributions and Paper Organization}
In Section \ref{sec:secretBlockSparsity}, we propose a novel communication protocol that leverages the advantageous recoverability of block-sparse signals in order ensure privacy. We provide the encoding and decoding strategies of Alice and Bob, respectively. In our design Alice transmits secretly to Bob a block structure and uses it to encode her message, which can be done at a very low transmission rate, while the channel matrix $\bA$ is \emph{not} left for Alice to design. To the authors knowledge, the showcased protocol is the first linear inverse problem based privacy method that does not require $\bA$ to be secretly shared. Furthermore, Corollary \ref{cor:singleSnapshotPrivacy} guarantee that Alice can adjust the block length and the sparsity level of the message she transmits so that the transmission is probably identifiable for Bob and unidentifiable in Eve's perspective as the signal length increases.

In Section \ref{sec:eavesdroppingHighOrderMoments}, we consider the possibility for Eve to recover the secret block structure from the observation of {\em multiple} snapshots of the observation $\{\by_\ell\}$ that have been generated by Alice with the same block structure $\cB$. We show in Corollary~\ref{cor:asymptoticVulnerability} that, depending on Alice's choice of the block length and sparsity level, it is possible to extract $\cB$ from the fourth order moments of the observation and propose an eavesdropping algorithm to that end. Finally, we investigate the case of a finite number of snapshots and derive an upper bound of the rate at which Alice must recycle $\cB$ to prevent Eve from deciphering Bob's messages. We present numerical results that validate our theoretical findings.

\subsection{Notation}
Vectors of $\bbR^n$ and matrices of $\bbR^{n_1 \times n_2}$ are denoted by boldface letters $\ba$ and capital boldface letters $\bA$ respectively. The matrix norms $\left\Vert \bM \right\Vert_2$ and $\left\Vert \bM \right\Vert_{\mathrm{max}}$ refer to the spectral norm and the maximal absolute value of the entries in $\bM$, respectively. The Hadamard product between two matrices $\bM_1$ and $\bM_2$ is denoted $\bM_1 \odot \bM_2$. We write by $\bI_n$ the identity matrix and by $\bJ_n$ the all-one matrix in appropriate dimension $n\times n$.  A block structure over $\bbR^N$ into $R$ blocks is described by a mapping  $\cB : [1,\dots,N] \to [1,\dots,R]$, and is associated with the indicator matrix of $\bB \in \bbR^{N \times N}$ defined by
\begin{equation}
	\bB_{i,j} = \begin{cases}
		1 & \textrm{if } \cB(i) = \cB(j) ; \\
		0 & \textrm{if } \cB(i) \neq \cB(j).
		\end{cases}
\end{equation}
We denote by $\bx[r]$ the subvector of $\bx$ with entries $x_\ell$ verifying $\cB(\ell) = r$.  The ``block-$\ell_0$-norm'' of a vector $\bx$ is defined as $\norm{\bx}_{\cB,0} = \sum_{r=1}^R \bm{1}_{\bx[r] \neq \bm{0}}$ and counts the number blocks in $\bx$ that are not exactly equal to $\bm{0}$. For two fonctions $f$ and $g$, we use the Landau notations $f = \Omega(g)$, $f = \mathcal{O}(g)$ and $f = o(g)$ to denote that the ratio $\frac{f}{g}$ is asymptotically greater than some constant $C$, smaller than $C$, and tends to $0$, respectively.

\section{Privacy with Block Sparsity}\label{sec:secretBlockSparsity}

\subsection{Alice's encoding}\label{ssec:AliceEncoding}
 Alice constructs her message $\bx$ has follows. Given the knowledge of the channel matrix, Alice initializes the communication by drawing a block structure $\cB : [1,\dots,N] \to [1,\dots,R]$ and sends this structure to Bob over the secret channel. We highlight that this exchange only requires $R \log_2(N)$ bits of information which significantly less than schemes relying on exchanging the  matrix $\bA$. For simplicity, we assume that the $R$ blocks have equal block size $d$, {\em i.e.} $N = Rd$. Next, Alice selects a probability of block activation $p \in [0,1]$, and encodes her message in a block-sparse vector $\bx$. In the sequel, we assume that $\bx$ is distributed according to a block Bernouilli-Gaussian distribution such that
\begin{equation}\label{eq:blockBernouilliGaussian}
	\bx[r] = \begin{cases}
		\bm{0}_d   & \textrm{w.p. } 1-p \\
		\bz[r] & \textrm{w.p. } p,
	\end{cases}
\end{equation}
where $\bz[r] \sim \cN(\bm{0}_d, \bI_d) $ is a random i.i.d. standard Gaussian vector of dimension $d$.

\subsection{Bob's decoding} \label{ssec:bobDecoding}
At the public channel output, Bob receives a vector $\by = \bA \bx$, and leverages $\cB$ that was securely sent by Alice to recover the ground truth message $\bx$. To do so, Bob formulates the block compressed sensing problem
\begin{align}\label{eq:bob_BlockCS}
	\hat{\bx}_{B} = \argmin_{\bx\in\bbR^N} \norm{\bx}_{\cB,0} \textrm{ such that } \by = \bA \bx.
\end{align}
Harnessing a block-sparse prior in compressed sensing has been extensively shown in the literature to enhance the identifiability of \eqref{eq:bob_BlockCS} and to allow an exact reconstruction of the message with much fewer measurements than classical compressed sensing \cite{eldar2009robust,baraniuk2010model, gribonval2003sparse}. However, directly solving \eqref{eq:bob_BlockCS} remains NP-hard in the generic case, due to the combinatorics inherent to the minimization of $\norm{\bx}_{\cB,0}$. Thus, Bob computes instead an estimate of $\widehat{\bx}_{B,\ell}$ using a polynomial time algorithm of his choice. Among the many addressed in the literature, Block Matching Pursuit (Block MP) \cite{bach2008consistency}, Block Basis Pursuit (Block BP) \cite{eldar2010block} or Block Iterative Harding Thresholding (Block IHT) \cite{baraniuk2010model}, have been proposed with provable performance guarantees. Proposition \ref{prop:successBob} recalls of the block length is large enough and the number of observation per non-zero elements remains constant then Bob can provably recover $\bx$.

\begin{proposition}[Success of Bob's decoding \cite{baraniuk2010model}]\label{prop:successBob}
	Suppose that $\bA$ is a matrix with i.i.d. random Gaussian entries. If
	\begin{align}\label{eq:BobSuccessRegime}
	    \log(p^{-1})  ={}& o(\log(d)); &
	    \frac{M}{pN} ={}& \Omega(1)
	\end{align}
	in the limit where $N \to \infty$, then Bob can recover $\bx$ asymptotically almost surely.
\end{proposition}

\subsection{Privacy Guarantees under a Single Snapshot}\label{ssec:privacySingleSnapshot}

If only one snapshot $\by$ is observed, it is impossible for Eve to reliably infer $\cB$, which remains ambiguous even with the perfect knowledge of $\bx$. Therefore, under her perspective, the best possible approach consists in attempting to recover $\bx$ without leveraging the existence of a latent block structure in the message. This amounts to solving the classical compressed sensing program
\begin{equation}\label{eq:eve_OptimalEstimator}
	\hat{\bx}_E = \argmin_{\bx\in\bbR^N} \norm{\bx}_{0} \textrm{ such that } \by = \bA \bx.
\end{equation}
The identifiability condition $\bx = \hat{\bx}_E$ of \eqref{eq:eve_OptimalEstimator} is well-understood to be related to the Restricted Isometry Property (RIP) of the measurement operator \cite{candes2008restricted}. In the case of a Gaussian matrix $\bA$, the following proposition, links the asymptotic failure of \eqref{eq:eve_OptimalEstimator} as a function of the parameters of the model.

\begin{proposition}[Failure of Eve's decoding \cite{blanchard2011compressed}]\label{prop:failureEve}
	Suppose that $\bA$ is a matrix with i.i.d. random Gaussian entries. Then if
	\begin{align}\label{eq:failureEveCondition}
		\frac{M}{pN} ={}& o\left( -\log(p)^{-1} \right)
	\end{align}
	holds in the limit where $N \to \infty$, then the solution of algorithm \eqref{eq:eve_OptimalEstimator} is different from $\bx$ with overwhelming probability.
\end{proposition}

Altogether, Propositions \ref{prop:successBob} and \ref{prop:failureEve} suggest that, given the dimensions $M$ and $N$ of $\bA$, Alice can select the parameters $p$ and $d$ so that \eqref{eq:BobSuccessRegime} and \eqref{eq:failureEveCondition} are jointly satisfied.

\begin{corollary}[Single snapshot privacy]\label{cor:singleSnapshotPrivacy} If Alice selects the determinantal ratio\footnote{The parameter $\alpha$ must be greater than $1$ for the problem to be identifiable.} $\alpha \triangleq \frac{M}{pN} > 1$ with $d \geq \frac{M}{\alpha N}$, then the protocol is asymptotically private to the exchange of a single message in the limit $N \to +\infty$.
\end{corollary}

\begin{figure}[t]
	\centering
	\includegraphics[width=\columnwidth]{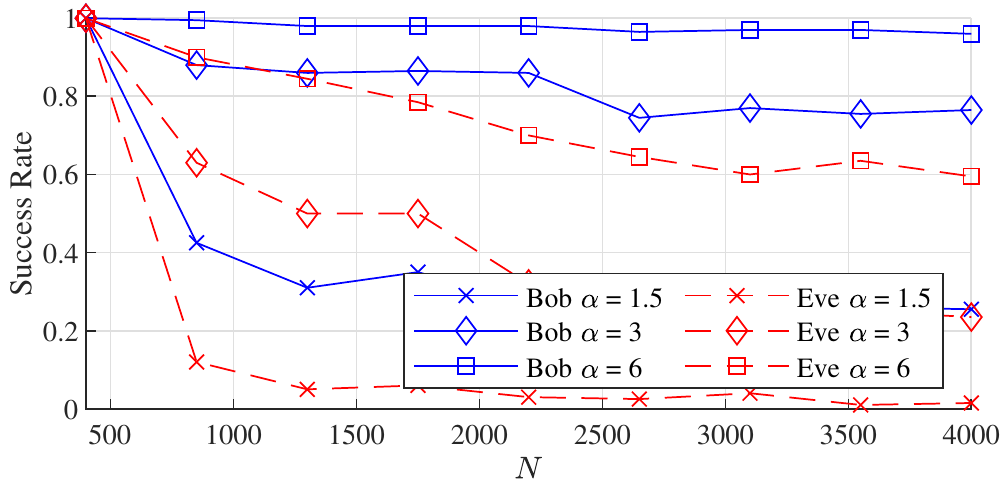}
	\caption{Success Rate of Bob and Eve to recover $\bx$ for different values of the ratio $\alpha = \frac{M}{pN}$. Here the parameters are set to $M = 200$, $d = 50$. The results are averaged over 200 trials.}
	\label{fig:singleSnapshotComparison}
\end{figure}

Fig.~\ref{fig:singleSnapshotComparison} shows the success rate of Bob and Eve to recover $\bx$ via the Block-BP and BP algorithms respectively, for different values of the ratio $\alpha$.

\section{Eavesdropping via Higher Order Moments}\label{sec:eavesdroppingHighOrderMoments}

\subsection{Structure of the Moments}\label{sec:perfectCovariance}
In order to reduce the usage of the secure channel, we want to understand the reusability of $\cB$ in transmitting several independent signals $\{\bx_1, \dots, \bx_L \}$. In that scenario, if Eve can acquires multiple snapshots of observation $\{\by_1, \dots, \by_L \}$ given by $\by_\ell = \bA \bx_\ell$, $\ell = 1,\dots,L$, and under the knowledge of the prior distribution \eqref{eq:blockBernouilliGaussian} of $\bx$, she can attempt to gain statistical information about $\cB$ without having to reconstruct the messages by studying the posterior distribution of $\by$. In particular, if the expectancy $\rE[\bx] = \bm{0}_N$ and covariance $\bSigma_{\bx} = p \bI_N$ of $\bx$ carry no information about the block structure $\cB$, the even forth order moments of $\bx$ given by
\begin{align}\label{eq:fourthOrderMomentOfX}
	\rE[x_\ell^2 x_{\ellp}^2] ={}& \begin{cases}
		3 p  & \text{if } \ell = \ellp \\
		p  & \text{if } \cB(\ell) = \cB(\ellp) \text{ and } \ell \neq \ellp \\
		p^2  & \text{if } \cB(\ell) \neq \cB(\ellp),
	\end{cases}
\end{align}
\emph{do} encode information about $\cB$. Additionally, as the odd fourth order moments of $\bx$ are equal to zero, Eve can restrict herself to the study of the covariance $\bSigma_\bv$ of the vector $\bv = \left( \bA^{\sT} \by \right) \odot \left( \bA^{\sT} \by \right)$. The following proposition indicates that the indicator matrix $\bB$ of the block structure $\cB$ can be approximately inferred from a proper translation of $\bSigma_\bv$.

\begin{proposition}[Structure of the 4th order moments]\label{prop:fourthOrderMoments}
	Denote by $\bSigma_\bv$ the covariance matrix of the vector $\bv$. Moreover let by $\bH = \beta_1 \bI_N + \beta_2 \bJ_N$ the matrix given by the coefficients
		\begin{align}\label{eq:valuesOfH}
			\beta_1 ={}& 8p - 2p^2, &
			\beta_2 ={}& 2p^2,
		\end{align}
	there exists $C > 0$ such that the inequality
	\begin{align}\label{eq:boundOnCovariance}
		\left\Vert \frac{1}{3p(1-p)} \left(\bSigma_\bv - \bH \right) - \bB \right\Vert_{\max}
		\leq{} &  \frac{d N \log^2(N)}{M^2}
	\end{align}
	holds with pr. greater than $1 - CN^{-1}$ when $N \to \infty$.
\end{proposition}
A sketch proof of Proposition \ref{prop:fourthOrderMoments} is presented in Appendix \ref{sec:proofOfFourthOrderMoments}. If the parameter $p$ and $d$ selected by Alice were known to Eve, she could compute $\bH$ and use the matrix $\tilde{\bB} = \left(3p(1-p)\right)^{-1} \left(\bSigma_\bv - \bH \right)$ as a first estimate of $\bB$. Additionally, since $\bB$ takes binary values, the matrix $\hat{\bB}$ obtained by rounding each entry of $\tilde{\bB}$ to its closest value in $\{0,1\}$ is equal to $\bB$ whenever the right hand side of \eqref{eq:boundOnCovariance} is smaller that $1/2$, the messages $\{\bx_\ell\}$ can subsequently be recovered by Eve, who solves \eqref{eq:bob_BlockCS} with its estimate $\hat{\cB}$.
\begin{corollary}[Asymptotic vulnerability]\label{cor:asymptoticVulnerability}
If $d = \mathcal{O} \left( \frac{N \log^2(N)}{M^2} \right) $, then the structure $\cB$ and the messages $\{\bx_\ell\}$ are asymptotically identifiable from the fourth order moment $\bSigma_\bv$.
\end{corollary}

\subsection{Estimation with a Finite Number of Snapshots}

\begin{algorithm}[t]
  \caption{Eavesdropping of the Block Structure
    \label{alg:eavesdroppingOfTheBlocks}}
  \begin{algorithmic}[1]
    \Function{EstimateBlocks}{$\bY, \bA, p, d$}
      \Let{$\bV$}{$\left(\bA^\sT  \bY\right) \odot \left(\bA^\sT  \bY\right)$}
      \Let{$\rE[\bv]$}{$ p \diag((\bA^\sT \bA)^2) $}
			\Let{$\hat{\bSigma}_\bv$} {$\frac{1}{L} \sum_{\ell=1}^L \left(\bv_\ell - \rE[\bv] \right) \left(\bv_\ell - \rE[\bv] \right)^{\sT}$}
			\Let{$\bH$}{$\beta_1 \bI_N + \beta_2 \bJ_N$}  \Comment{With $\left( \beta_1,\beta_2 \right)$ as given in \eqref{eq:valuesOfH}}
			\Let{$\tilde{\bB}$}{$p^{-1}{(1-p)}^{-1}\left(\hat{\bSigma}_\bv - \bH\right)$}
			\Let{$\hat{\bB}$} {$\bm{1}_{\tilde{\bB} > \frac{1}{2}}$} \Comment{Rounding operation}

      \State \Return{$\hat{\bB}$}
    \EndFunction
  \end{algorithmic}
\end{algorithm}

In practice, Eve has access to a limited number of snapshots $L$ before Alice terminates the communication or refreshes the structure $\cB$. Consequently, the true covariance $\bSigma_\bv$ always remains unknown to Eve. Instead, she can attempt to estimate $\cB$ from the empirical estimator of the covariance given by
\(
\hat{\bSigma}_{\bv} = \frac{1}{L} \sum_{\ell=1}^L \left(\bv_\ell - \rE[\bv] \right) \left(\bv_\ell - \rE[\bv] \right)^{\sT},
\)
where $\rE[\bv] = p \diag \left ( \left(\bA^{\sT} \bA\right)^2 \right)$ and $\diag(\cdot)$ is the operator that stacks the diagonal elements of a $N \times N$ matrix into an $N$-dimensional vector.

As for the conclusion drawn in Subsection \ref{sec:perfectCovariance}, Eve can compute the estimator $\hat{\bB}$ by rounding the matrix $\widehat{\bB} =  \left(3p(1-p)\right)^{-1} \left(\hat{\bSigma}_\bv - \bH \right)$. This complete procedure is summarized in Algorithm \ref{alg:eavesdroppingOfTheBlocks}.  However, the quality of the estimate $\tilde{\bB}$ will be worsened by the estimation error on $\bSigma_\bv$. The next  lemma provide a high probability bound on this error via the matrix Bernstein inequality (e.g.~ \cite{tropp2015introduction}).

\begin{lemma}[Concentration of $\hat{\bSigma}_\bv$]\label{lem:concentrationOfTheCovariance}
There exists two constants $C_1, C_2$ such that the inequality
\begin{equation}\label{eq:covarianceConcentration}
	\left\Vert \hat{\bSigma}_\bv - \bSigma_\bv \right\Vert_2 \leq \frac{p N^2 \log^2(N) \log(L)}{M^2 \sqrt{d}\sqrt{L}}
\end{equation}
holds with probability greater than $1 - C_1 N^{-1} - C_2 L^{-1}$.
\end{lemma}
Equations \eqref{eq:boundOnCovariance} and \eqref{eq:covarianceConcentration} with the bound $\left\Vert \cdot \right\Vert_{\mathrm{max}} \leq \left\Vert \cdot \right\Vert_2$ yields an upper bound on the number of snapshots that are necessary of Eve to compromise the protocol.

\begin{corollary}[Vulnerability under a finite number of snapshots]\label{cor:vulnerabilityFiniteSnapshots}
Let $L_{\mathrm{crit}} \triangleq \frac{N^4 \log^4(N)}{M^4 d}$. If $d = \mathcal{O} \left( \frac{N \log^2(N)}{M^2} \right)$ and $L = \Omega\left(L_{\mathrm{crit}} \right)$ then Algorithm  \ref{alg:eavesdroppingOfTheBlocks} recovers the truth structure $\cB$ with overwhelming probability when $N \to \infty$.
\end{corollary}

\begin{figure}[t]
	\centering
	\includegraphics[width=\columnwidth]{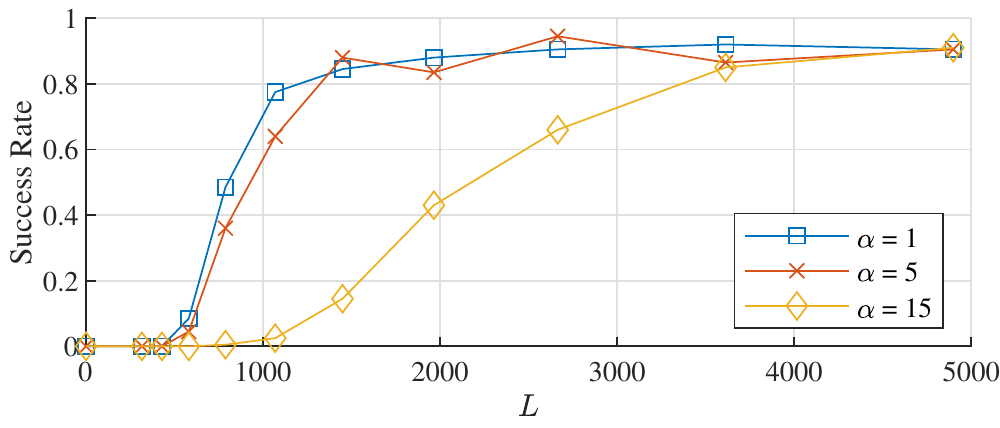}
	\caption{Success rates of Algorithm \ref{alg:eavesdroppingOfTheBlocks} to recover the block structure $\cB$ as a function of the number of snapshots $L$ and different ratios \mbox{$\alpha = \frac{M}{pN}$}. Herein, $N = 1000$, $M = 400$ and $d=20$. Results are averaged over 200 trials. }
	\label{fig:methodOfMoments}
\end{figure}

If Alice follows the scaling $d \sim \frac{M}{\alpha N}$ proposed in Subsection \eqref{ssec:privacySingleSnapshot} the lifespan of $\cB$ will be $L_{\mathrm{max}} \sim \frac{\alpha N^5 \log^4(N)}{M^5}$ which suggests that: large channel are more robust to statistical eavesdropping; and that higher values of the determinantal parameter $\alpha$ increases the lifespan of $\cB$. This last observation is corroborated by  Fig.~\ref{fig:methodOfMoments}. However increasing $\alpha$ amounts to decreasing the amount of information transmitted in each message $\bx_\ell$. This last observation suggests the existence of a trade-off between the achievable communication rate and the robustness of the privacy of the proposed protocol which is proposed for future study.

\appendix

\section{Proof of Proposition 3}\label{sec:proofOfFourthOrderMoments}
Herein, we provide a proof sketch of Proposition \ref{prop:fourthOrderMoments}. First, let $\bM = \bA^{\sT} \bA$, and denote by $\bZ$:
\begin{multline}\label{eq:ZDefinition}
	\bZ = 2 \diag(\bM)\diag(\bM)^\sT \cdot \\
	\left(2p \bI_N + p\left(1 - p \right) \bB  + p^2 \bJ_N \right) \cdot \diag(\bM)\diag(\bM)^\sT
\end{multline}
with expected value $\rE_{\bM}[\bZ] = 4p \bI_N +2 p\left(1 - p \right) \bB  + 2p^2 \bJ_N$. A direct calculation yields the expression of the covariance~$\bSigma_\bv$,
\begin{equation}\label{covarianceOfSigmaV}
	\bSigma_\bv = p(1-p)\left( \bM \odot \bM \right) \left( 2\bI_N + \bB \right) \left( \bM \odot \bM \right) + \bZ.
\end{equation}
Next, as the entries of $\bA$ are i.i.d Gaussian with \mbox{$a_{i,j} \sim \cN\left(0, \frac{1}{M}\right)$} there exists the event
	\(\label{eq:conditionEpsilon}
		\max_{i,j} \left\vert \inp{\ba_i}{\ba_j}^2 - \delta_{i,j} - \frac{1}{M} \right\vert \leq \frac{\log(N)}{M}
	\)
	holds with probability greater than $1 - C_0 N^{-1}$ for some $C_0 > 0$,
which leads after some algebra to
\begin{subequations}\label{eq:boundProp2-1}
	\begin{align}
		\left\Vert \left(\bM \odot \bM \right)\bI_N \left(\bM \odot \bM \right) - \frac{N}{M^2} \bJ_N \right\Vert_\mathrm{max} \leq{}& \frac{N \log^2(N)}{M^2}, \\
		\left\Vert \left(\bM \odot \bM \right) \bB \left(\bM \odot \bM \right)  - \frac{dN}{M^2} \bJ_N \right\Vert_\mathrm{max} \leq{}& \frac{dN \log^2(N)}{M^2}.
	\end{align}
\end{subequations}

The next Lemma, obtained uniformly controlling each entries using Hanson-Wright inequality \cite{rudelson2013hanson, klochkov2020uniform}, proposes a bound on the quantity $ \bW = \bZ - \rE_{\bm{M}}[\bZ]$.
\begin{lemma}[Uniform Hanson-Wright type inequality]\label{lem:Hanson-Wright}
	There exists a constant $C > 0$ such that the inequality
	\(
		\left\Vert \bW \right\Vert_\mathrm{max} \leq \frac{p \sqrt{dN}{\log^6({N})}}{M^2}
	\)
	holds with probability greater than $1 - C N^{-1}$.
\end{lemma}
One concludes on the desired statement with Lemma \ref{lem:Hanson-Wright} by applying the triangle inequality on \eqref{covarianceOfSigmaV}, \eqref{eq:boundProp2-1} for a sufficiently large $N$. \qed

\printbibliography

\end{document}